\documentclass[twocolumn,showpacs,nofootinbib,aps,
prd]{revtex4}
\usepackage{graphicx,dcolumn,bm,amsfonts,amsmath,color,xcolor,amsthm}
\usepackage[colorlinks=true, citecolor=blue, linkcolor=blue,urlcolor=blue]{hyperref}
\usepackage{slashed}
\usepackage{ulem}
\usepackage{txfonts}
\DeclareMathAlphabet{\mathcal}{OMS}{cmsy}{m}{n}
\DeclareSymbolFont{largesymbols}{OMX}{cmex}{m}{n}

\allowdisplaybreaks

\begin{document}
\title{Non-perturbaitve effects for the isoscalar light vector $\omega$-meson in charmed meson semileptonic decays}
\author{Yin-Long Yang$^*$}
\author{Fang-Ping Peng\footnote{Yin-Long Yang and Fang-Ping Peng contributed equally to this work.}}
\author{Yan-Ting Yang}
\author{Hai-Bing Fu}
\author{Sheng-Quan Wang}
\email{sqwang@alu.cqu.edu.cn (corresponding author)}
\address{Department of Physics, Guizhou Minzu University, Guiyang 550025, P.R.China}

\begin{abstract}
Motivated by the renewed attention from the recent BESIII experiment on the semileptonic decay $D\to V \ell\nu_{\ell}$ (where $V$ denotes a vector meson), we investigate semileptonic decay $D^+\to \omega \ell^+\nu_{\ell}$ within the framework of QCD light-cone sum rule in this work. By constructing correlation function with right-handed chiral current, the transverse twist-2 light-cone distribution amplitudes (LCDA) $\phi^{\perp}_{2;\omega}(x,\mu)$ dominates the contribution in TFFs. We study the properties of twist-2 LCDA $\phi^{\perp}_{2;\omega}(x,\mu)$ through light-cone harmonic oscillator model. Applying it to the TFFs, we obtained $A_1(0)=0.537^{+0.053}_{-0.053}$, $A_2(0)=0.540^{+0.068}_{-0.068}$, $V(0)=0.754^{+0.079}_{-0.079}$, and $A_0(0)=0.553^{+0.044}_{-0.043}$ at large recoil point. Two TFF ratios are $r_V=1.40^{+0.21}_{-0.19}$ and $r_2=1.01^{+0.17}_{-0.16}$. After extrapolating those TFFs to the whole physical $q^2$ region by using the simplified $z(q^2,t)$ series expansion, the ratio of longitudinal and transverse decay widths is $\Gamma_{\rm{L}}/\Gamma_{\rm{T}}=0.987^{+0.107}_{-0.121}$. Then, we get branching fraction $\mathcal{B}(D^+\to \omega e^+\nu_e)=(1.848^{+0.365}_{-0.330})\times 10^{-3}$ and $\mathcal{B}(D^+\to \omega \mu^+\nu_{\mu})=(1.782^{+0.334}_{-0.303})\times 10^{-3}$, which is in good agreement with BESIII and CLEO Collaborations. Taking into account the secondary decay $\omega\to \pi^+\pi^-\pi^0$, we predict branching fraction of five body decay as $\mathcal{B}(D^+\to \omega(\to \pi^+\pi^-\pi^0)e^+\nu_{e})=(1.648^{+0.341}_{-0.305})\times 10^{-3}$ and $\mathcal{B}(D^+\to \omega(\to \pi^+\pi^-\pi^0)\mu^+\nu_{\mu})=(1.589^{+0.313}_{-0.281})\times 10^{-3}$. Finally, we predict the forward-backward asymmetry $A_{\rm{FB}}^{\ell}$, lepton-side convexity parameter $C^{\ell}_{\rm{F}}$, longitudinal (transverse) polarization $P_{\rm{L}(\rm{T})}^{\ell}$, as well as longitudinal polarization fraction $F_{\rm{L}}^{\ell}$.
\end{abstract}
\maketitle

\section{Introduction}\label{Sec:I}
Charmed hadron decays and precision tests of the Standard Model (SM) have long been key research fields for the BESIII Collaboration. It relies on $e^+e^-$ collisions produced by the double-ring collider BEPCII, which is working at the center-of-mass energy range from 1.85 to 4.95 $\rm{GeV}$. This year, BEPCII has been successfully upgraded again, achieving a high peak luminosity at a beam energy of $2.35~\rm{GeV}$. Its peak luminosity and daily integrated luminosity reached $7.35\times 10^{32}~\rm{cm}^ {-2}s^{-1}$ and $35.1~\rm{pb}^{-1}$, respectively. The peak luminosity and integral luminosity have more than doubled compared to pre-upgrade levels. This marks a significant leap in the core performance of the collider. The decay channels of charmed hadrons are abundant. Among them, semileptonic decays of charmed mesons is an important branch. In contrast to nonleptonic (hadronic) weak decays, these processes involve no significant final-state strong interactions or interference among multiple amplitudes, making them theoretically cleaner. These decays depend on the Cabibbo-Kobayashi-Maskawa (CKM) matrix elements $|V_{cs}|$ and $|V_{cd}|$, which describe the flavor-changing transitions among quarks. Meanwhile, the strong interaction binding effects are confined to the hadronic current and parameterized as transition form factors (TFFs) in semileptonic decay process, making it a clean channel for determining the CKM matrix elements. In this context, the BESIII Collaboration will continue to provide more precise measurements of observables in relevant semileptonic decay processes.

Among the various semileptonic decay channels of $D$-mesons, the $D\to P\ell\nu$ decays (where $P$ denotes a pseudoscalar meson) are theoretically and experimentally more mature compared to those with scalar meson ($S$) or vector meson $(V)$ in the final state. In the field of experimental research, the BESIII collaboration has not only provided branching fraction measurements but has also precisely determined the differential decay widths across different $q^2$ region. Furthermore, it has also made precise predictions of TFFs, such as $D_{(s)}\to (\pi, K)\ell\nu$ decays~\cite{BESIII:2018xre,BESIII:2015tql}. For the $D\to S\ell\nu$ decay, both experimental and theoretical data are relatively scarce. The reason is that the internal structure of the scalar family has various possible assumptions~\cite{Cheng:2005nb,Jaffe:1976ig,Weinstein:1990gu,Weinstein:1982gc,Weinstein:1983gd}, which makes its research background more complicated. In contrast, the vector mesons such as $\rho, \phi, K^*$, and $\omega$ have well-established composition of $q\bar{q}$ state. In recent years, the BESIII collaboration has already updated the observables for the semileptonic decays $D_{(s)}\to (\rho, \phi, K^*) \ell\nu_{\ell}$~\cite{BESIII:2024lnh,BESIII:2024mot,BESIII:2023opt}. For semileptonic decay $D^+\to \omega \ell^+\nu_{\ell}$, no new experimental data have been reported and it has only been discussed twice in the past. In 2015, the BESIII Collaboration reported $\mathcal{B}(D^+\to \omega e^+\nu_{e})=(1.63\pm0.11\pm0.08)\times 10^{-3}$ and provided the ratio of hadronic TFF at zero momentum transfer for the first time, $i.e.,$ $r_V=1.24\pm0.09\pm0.06$ and $r_2=1.06\pm0.15\pm0.05$~\cite{BESIII:2015kin}. And $\mathcal{B}(D^+\to \omega \mu^+\nu_{\mu})=(1.77\pm0.18\pm0.11)\times 10^{-3}$ is measured in 2020~\cite{Ablikim:2020tmg}. The $\omega$-meson and the $\rho$-meson share several similar properties. For example, they are vector mesons with the same $P$-parity and $C$-parity, and their masses are very close. Both of them contain $u$ and $d$-quarks, which are one of the golden decay channels that extract the CKM matrix element $|V_{cd}|$ via $c\to d \ell\nu_{\ell}$ process. Currently, in both experimental and theoretical aspects, the discussion of physical quantities for $D\to \rho \ell\nu_{\ell}$ process is richer than that for $D\to \omega \ell\nu_{\ell}$ process. But under the same circumstances, the physical background of $D\to \omega \ell\nu_{\ell}$ process may be more concise, and the physical quantities will be more accurate at present. The reason is that previous works on the decay width of $D\to \rho \ell\nu_{\ell}$ process often adopted the expression under the narrow-width approximation, $i.e.$, the ideal case where the width of $\rho$-meson tends to zero~\cite{Cheng:2017pcq,Ball:1993tp,Zhong:2023cyc,Bernard:1991bz,Wang:2002zba}. According to PDG~\cite{ParticleDataGroup:2024cfk}, its width is $\Gamma_{\rho}=147.4\pm0.8~\rm{MeV}$, which constitutes about $19\%$ of its own mass. This indicates that $\rho$-meson cannot be regarded as a stable particle and can easily decay into $\pi^+\pi^-$. Therefore, we cannot ignore the effects arising from its width in actual theoretical calculations. In contrast, the width of the $\omega$-meson is $\Gamma_{\omega}=8.68\pm0.13~\rm{MeV}$, which is only about $1\%$ of its mass. This indicates that we can approximately treat it as a stable particle with a well determined mass, and its width effects can be safely neglected. Therefore, in $D$-meson decays, the $\omega$-meson is a more suitable candidate for precision studies than the $\rho$-meson.

Furthermore, there exists an interesting phenomenon between these two mesons. The neutral vector meson $\omega$ is characterized by an isospin of $I=0$. Its branching fraction of dominant decay channel is $\mathcal{B}(\omega\to \pi^+\pi^-\pi^0)= 0.892\pm0.007$. At the hadronic level, there exists a mixing effect between the $\rho^0$ and $\omega$-mesons. In the $\rho$-$\omega$ resonance region, the cross-section of $e^+e^-\to \pi^+\pi^-$ process exhibits a narrow interference shoulder, which arises from the superposition of the narrow $\omega$ resonance and broad $\rho$ resonance exchange amplitudes~\cite{Barkov:1985ac}. In fact, in the process of strong interaction participation, $\omega$-meson is not allowed to decay into pion pairs, because it violates the $G$-parity. However, the participation of the electromagnetic interaction and the isospin breaking caused by the mass difference of the $u,d$-quark will cause the pure isospin states $\rho_{I}$ (pure isovector $\rho$) and $\omega_{I}$ (pure isoscalar $\omega$) to mix, resulting in the mass eigenstates $\rho$ and $\omega$ being the superposition of two initial fields~\cite{Gao:1998gr,OConnell:1995nse,Maltman:1996kj,Glashow:1961dn}. The $\rho$-$\omega$ mixing can be described by the physical basis
\begin{align}
&\rho=\rho^{(I=1)}+\epsilon\omega^{(I=0)}, &&\omega=\omega^{(I=0)}-\epsilon\rho^{(I=1)}
\end{align}
where the complex mixing parameter $\epsilon$ is $-0.006 + 0.036i$~\cite{Urech:1995ry,OConnell:1995fwv}. This allows the decay channel $\omega\to \pi^+\pi^-$ to exist, with a branching fraction of $\mathcal{B}(\omega\to \pi^+\pi^-) =0.0153\pm0.0012$. The dominant decay channel of the $\rho$-meson is $\rho\to \pi\pi$ with a branching fraction of $\mathcal{B}(\rho\to \pi\pi)\backsim 1$. Experimentally, the $\rho$-meson is reconstructed via the kinematics of the two pions, allowing one to investigate some properties of $D\to\rho\ell\nu_{\ell}$ decays. The Belle Collaboration reconstructed the $\rho$-meson using $\pi^+\pi^-$ and further studied the $B\to \rho \ell\nu_{\ell}$ process, treating the $\omega$ as a fixed, small background component, and assigning a $100\%$ uncertainty to its size to cover possible contamination due to $\rho$-$\omega$ mixing or model inaccuracies~\cite{Belle:2013hlo}. Other multibody decays suffer from the same issue, such as $\omega \to \pi^+\pi^-, \omega\to \pi^0\pi^0\gamma$, and $B^{\pm}\to \rho^0\pi^{\pm}$~\cite{Guo:2000uc,Guetta:2000ra,Williams:1997nj}. In the past, they were usually considered as rare isospin-violating decays with tiny branching fractions. Thus, it was treated as negligible backgrounds. However, the $\rho$-$\omega$ mixing effect may need to be discussed more thoroughly with the ever-improving experimental precision in the further. In contrast, the semileptonic decay $D^+\to \omega\ell^+\nu_{\ell}$ has good advantages. Experimentally, the $\omega$-meson is typically reconstructed via $\pi^+\pi^-\pi^0$ to study the physical quantities of semileptonic decay~\cite{BaBar:2012dvs,BaBar:2012thb}. This approach avoids the $\rho$-$\omega$ mixing problem, which leads to a cleaner theoretical background and more controllable systematic uncertainties. Therefore, the semileptonic decay $D^+\to \omega\ell^+\nu_{\ell}$ is expected to be re-examined in the future experiments. Not only are the corresponding TFFs very likely to be experimentally determined, but some observables that are highly sensitive to new physics may also be measured, such as forward-backward asymmetry $A_{\rm{FB}}^{\ell}$, lepton-side convexity parameter $C^{\ell}_{\rm{F}}$, etc. In this work, we try to provide a reliable theoretical prediction for the physical quantities in semileptonic decay $D^+\to \omega\ell^+\nu_{\ell}$.

The hadronic current involved in the semileptonic decay can be parameterized into TFFs, which is the most important research object in theory. For $D^+\to \omega \ell^+\nu_{\ell}$, it can be described by four TFFs: $A_1(q^2), A_2(q^2), A_0(q^2)$ and $V(q^2)$. Currently, it has been studied by various nonperturbative methods, such as heavy quark effective field theory (HQEFT)~\cite{Wu:2006rd}, heavy meson and chiral symmetries (HM$\chi$T)~\cite{Fajfer:2005ug}, light-front quark model (LFQM)~\cite{Verma:2011yw}, relativistic quark model (RQM)~\cite{Faustov:2019mqr}, covariant confining quark model (CCQM)~\cite{Ivanov:2019nqd}. Compared with $D_{(s)}\to (\rho,\phi,K^*)$, the current theoretical group has relatively less discussion on the TFF of $D^+\to \omega$, and there is a lack of lattice QCD (LQCD) data. In the full physical region, LQCD can provide an accurate prediction of TFFs in the high $q^2$ region, while in the low and intermediate regions, the light-cone sum rule (LCSR) can offer precise predictions. The two can complement each other. So far, the LCSR method has successfully applied to heavy to light process, which allows a systematic inclusion of both hard scattering effects and soft contributions~\cite{Ball:2001fp,Khodjamirian:2000ds,Aliev:1995zlh,Khodjamirian:2002pk}. Compared with the traditional QCDSR, the difference of LCSR is that nonperturbative effects are no longer represented by vacuum condensates of different dimensions, but parameterized as light cone distribution amplitudes (LCDAs). Generally, when discussing the LCDAs of pseudoscalar or scalar mesons, we can uniformly distinguish the contributions of LCDAs according to the level of twists. However the LCDAs of vector mesons cannot be treated in this way. The reason is that in the operator product expansion (OPE) step, we encounter LCDAs of different polarization states arising from chiral-odd and chiral-even operators~\cite{Ball:1998sk}. Here, the longitudinal state and transverse are denoted as $\|$ and $\perp$, respectively. This situation results in the final analytical expressions of TFFs containing combinations of different LCDAs, rather than a single form. In general, there are fifteen LCDAs for vector mesons. When constructing the correlation functions, their contributions must be taken into account when using the traditional currents. Although twist-2 LCDAs dominate the contributions in LCSR, there is currently no in-depth discussion on the high twists, and the corresponding uncertainties introduced are relatively large. To address this, we can consider adopting appropriate methods. For example, by adopting the left-handed chiral current, only the chiral-even LCDAs contribute to OPE, and the longitudinal twist-2 LCDAs dominate the contribution. We have recently discussed the $B^+\to \omega$ using this approach, but one of the TFFs shows a sharp rising trend in the high $q^2$ region~\cite{Yang:2025gcm}. For the $B^+\to \omega$ process with a large momentum transfer range, this situation is still within a reasonable range, but we still expect a more stable TFF. However, in the $D^+\to \omega$, we can anticipate that this phenomenon may be more significant due to the smaller range of momentum transfer. Therefore, in this work, we no longer use the same method for calculation. Instead, we propose to construct the correlation function using the right-handed chiral current, which provides a new perspective for examining the behavior of TFFs. Under this method, only chiral-odd LCDAs will contribute, among which the transverse twist-2 LCDA $\phi^{\perp}_{2;\omega}(x,\mu)$ making the dominant contribution. Therefore, a reliable transverse twist-2 LCDA behavior is important for ensuring the accuracy of LCSR method and also helping us understand the momentum fraction distributions of partons inside $\omega$-meson in a particular Fock state.

In fact, there has been much more discussion on the longitudinal twist-2 LCDA than on the transverse twist-2 LCDA of vector mesons~\cite{Bakulev:1998pf,Stefanis:2015qha,Gao:2014bca,Forshaw:2012im,Almeida-Zamora:2023rwg,Ball:2007zt}. In particular, lattice QCD (LQCD) has also provided theoretical predictions for $\phi_{2;\rho,\phi,K^*}^{\|/\perp}(x,\mu)$~\cite{Braun:2016wnx, Segovia:2013eca, Hua:2020gnw}. In contrast, the relevant theoretical discussions on transverse twist-2 LCDA $\phi^{\perp}_{2;\omega}(x,\mu)$ of $\omega$-meson are very scare. Therefore, it is highly necessary for us to conduct a detailed study on the precise behavior of twist-2 LCDAs. For transverse twist-2 LCDA $\phi^{\perp}_{2;\omega}(x,\mu)$, it is commonly to use the Gegenbauer polynomial form based on conformal symmetry, where the Gegenbauer moment $a^{\perp}_{n;\omega}(x,\mu)$ contains nonperturbative information. Usually, one only focus on the discussion of the first order or the first two orders $a^{\perp}_{n;\omega}(x,\mu)$. Because the uncertainty of the current theoretical method for calculating the high-order $a^{\perp}_{n;\omega}(x,\mu)$ is very large, and it also leads to false oscillation in LCDA~\cite{Li:2022qul,Ball:2004ye}. For $\omega$-meson, the odd Gegenbauer moments vanish due to isospin symmetry, and only the even Gegenbauer moments remain. The first Gegenbauer moment $a^{\bot}_{2;\omega}(x,\mu)$ is first investigated in the late 20th century. When neglecting $u,d$ quark masses and $\rho$-$\omega$ mixing effects, and assuming the twist-2 LCDAs of $\rho$ and $\omega$-mesons are equal under proper normalized currents, $a^{\bot}_{2;\omega}(\mu_0)=0.2\pm0.1$ was obtained by QCDSR at the initial scale $\mu_0=1~\rm{GeV}$~\cite{Ball:1998sk}. Meanwhile, M. Dimoul and J. Lyon (DL) predict $a^{\bot}_{2;\omega}(\mu_0)=0.14\pm0.12$ in 2013~\cite{Dimou:2012un}, which was obtained by considering the value of $a^{\bot}_{2;\rho}(\mu_0)$ from both RBC and UKQCD Collaborations~\cite{Arthur:2010xf}, and QCDSR~\cite{Ball:2007rt}, with the original uncertainties conservatively doubled. Furthermore, the LCDA $\phi^{\perp}_{2;\omega}(x,\mu)$ is related to light cone wave function (LCWF), which is the integral projection of LCWF in the transverse momentum space. At present, it is difficult to obtain the specific form of LCWF from the first principle of QCD. A common approach is to construct phenomenological models of LCWF~\cite{Chang:2013pq,Bondar:2004sv,Brodsky:2006uqa,Ma:2004qf,Wu:2007rt,Huang:2004su}. In this work, we will construct a light-cone harmonic oscillator (LCHO) model based on Brodsky-Huang-Lepage (BHL) description, where the LCWF in the infinite momentum frame is related to equal-time WF in the rest frame~\cite{Fu:2014cna,Zhong:2011rg}. This phenomenological model has been successfully applied to various light mesons~\cite{Yang:2024ang, Hu:2024tmc, Hu:2023pdl, Wu:2022qqx, Fu:2016yzx, Yang:2024jlz, Wu:2025kdc}, which provides a reliable and specific analytical expression of LCWF.

This paper is organized as follows. In Section~\ref{Sec:II}, we present the calculation of the $D^+\to \omega$ TFFs using the LCSR method, the construction of the LCHO model for the twist-2 LCDA of $\omega$-meson, and the determination of the model parameters. In Section~\ref{Sec:III}, we show the detailed numerical analysis and discussion. Section~\ref{sec:IV} is used to be a summary.

\section{Theoretical Framework}\label{Sec:II}
For $D^+\to \omega\ell^+\nu_{\ell}$, it can be described by $c\to dW^{*+}$ and $W^{*+} \to \ell^+\nu$ in the tree diagram of SM. As we all know, leptons do not participate in the strong interaction, so that weak interactions and strong interactions can be separated when calculating decay amplitude $\mathcal{M} (D^+\to \omega\ell^+\nu)$. With the effective Hamiltonian for $c\to d$ transition, we have
\begin{align}
\mathcal{M} (D^+\to \omega\ell^+\nu)=\frac{G_F}{\sqrt{2}}V_{cd}H^{\mu}L_{\mu},
\end{align}
where $G_F=1.166\times 10^{-5}~\rm{GeV}^{-2}$ is Fermi coupling constant, $L_{\mu}=\bar{\nu}_{\ell}\gamma _{\mu}(1-\gamma _5)\ell^+$ is leptonic current, and $ H^{\mu}=\langle \omega|V^{\mu}-A^{\mu}|D^+ \rangle$ is hadron matrix element with flavor-changing vector current $H^{\mu}=\bar{q}\gamma^{\mu}c$ and axial-vector currents $A^{\mu}=\bar{q}\gamma^{\mu}\gamma_5c$.  When calculating the squared decay amplitude $|\mathcal{M}|^2$, $L_{\mu}$ is relatively simple, as it does not involve any nonperturbative strong interaction parameters and only requires summing over all possible lepton spins. $H^{\mu}$ contains all the strong interaction information, which is a very complex computational object. Fortunately, we can use Lorentz covariance to parameterize it into universal TFFs,
\begin{align}
&\langle \omega(p,\lambda)|\bar{d}\gamma_{\mu}(1-\gamma_5)c|D^+(p+q) \rangle\nonumber\\
= & -i e_{\mu}^{*(\lambda)}(m_{D^+}+m_{\omega}) A_{1}(q^2) \nonumber\\
& +i(e^{*(\lambda)} \cdot q) \frac{A_{2}(q^2)(2 p+q)_{\mu}}{m_{D^+}+m_{\omega}}\nonumber\\
& +i q_{\mu}(e^{*(\lambda)} \cdot q) \frac{2 m_{\omega}}{q^{2}}[A_{3}(q^2)-A_{0}(q^2)]\nonumber\\
& +\epsilon_{\mu \nu \alpha \beta} e^{*(\lambda) \nu} q^{\alpha} p^{\beta} \frac{2 V(q^2)}{m_{D^+}+m_{\omega}}.
\label{Eq:matrix difen}
\end{align}
Then, the differential decay width over $q^2$ and $\cos \theta$ can be obtained~\cite{Ivanov:2015tru}
\begin{align}
\hspace{-0.5cm}\frac{d\Gamma (D^+\to \omega\ell ^+\nu _{\ell})}{dq^2 d\cos \theta}&=\frac{G_{F}^{2}}{(2\pi )^3}|V_{cd}|^2\frac{\lambda ^{1/2}(q^2-m_{\ell}^{2})^2}{64M_{D^+}^{3}q^2}  \nonumber\\
&\times \biggl[ (1+\cos ^2\theta )\mathcal{H} _U+2\sin ^2\theta \mathcal{H} _L \nonumber\\
&+2\cos \theta \mathcal{H} _P +\frac{m_{\ell}^{2}}{q^2}(\sin ^2\theta \mathcal{H} _U \nonumber\\
 &+2\cos ^2\theta \mathcal{H} _L+2\mathcal{H} _S-4\cos \theta \mathcal{H} _{SL})\bigg] .
\end{align}
where $\lambda\equiv \lambda(m^2_{D^+},m^2_{\omega},q^2)=m^4_{D^+}+m^4_{\omega}+q^4-2(m^2_{D^+}m^2_{\omega}+m^2_{\omega}q^2+m^2_{D^+}q^2)$. For the convenience of representation, the helicity amplitudes $\mathcal{H}_i$ are introduced, which are related to four universal TFFs~\cite{Ivanov:2019nqd}
\begin{align}
&\mathcal{H} _U=|H_+|^2+|H_-|^2, && \mathcal{H} _L=|H_0|^2,\nonumber\\
&\mathcal{H} _P=|H_+|^2-|H_-|^2,&&\mathcal{H} _S=|H_t|^2,\nonumber\\
&\mathcal{H} _{SL}=\mathrm{Re} (H_0H_{t}^{\dagger}),
\end{align}
with
\begin{align}
H_\pm(q^2)
&= \frac{\lambda^{1/2}}{m_{D^+}+m_\omega}
\bigg[ \frac{(m_{D^+}+ m_\omega)^2}{ \lambda^{1/2} }A_1(q^2) \mp   V(q^2) \bigg],
\nonumber\\
H_0(q^2)
&=\frac{1}{2m_\omega\sqrt{q^2}}\bigg[ (m_{D^+}+m_\omega)(m_{D^+}^2\!\!-\!m_\omega^2\!-\!q^2)A_1(q^2)
\nonumber\\
&\hspace{0.4cm}-\frac{\lambda}{m_{D^+}+m_\omega}A_2(q^2) \bigg],
\nonumber\\
H_t(q^2)
&=\frac{\lambda^{1/2}}{\sqrt{q^2}}A_0(q^2).
\end{align}
Typically, when the lepton mass $m_{\ell}\to 0$, the vector TFF $A_0(q^2)$ makes no contribution to the differential decay width. However, when aiming to study the lepton mass effects, $A_0(q^2)$ needs to be retained. Meanwhile, we can also define some other physical observables that are equally sensitive to new physics, such as the forward-backward asymmetry $A_{\rm{FB}}^{\ell}$, lepton-side convexity parameter $C^{\ell}_{\rm{F}}$, longitudinal (transverse) polarization $P_{\rm{L}(\rm{T})}^{\ell}$  of the charged lepton in the final state, as well as longitudinal polarization fraction $F_{\rm{L}}^{\ell}$ of final vector meson $\omega$. Their specific expressions are defined as~\cite{Gutsche:2015mxa}
\begin{align}
A_{\mathrm{FB}}^\ell(q^2)
&=\frac{\int_0^1 \cos\theta d\Gamma/ d\cos\theta-\int_{-1}^0 d\cos\theta d\Gamma/d\cos\theta}{\int_0^1d\cos\theta d\Gamma/d\cos\theta+\int_{-1}^0d\cos\theta d\Gamma/d\cos\theta}
\label{Eq:AFB}
\nonumber\\
&=\frac{3}{4}\frac{\mathcal{H}_P-2\frac{m^2_{\ell}}{q^2}\mathcal{H}_{SL}}{\mathcal{H}_
{\mathrm{total}}},
\\
C_\mathrm{F}^\ell(q^2)
&=\frac{3}{4}\left(1-\frac{m^2_{\ell}}{q^2}\right)\frac{\mathcal{H}_{U}-2\mathcal{H}_{L}}
{\mathcal{H}_\mathrm{total}},
\\
P_\mathrm{L}^\ell(q^2)
&=\frac{(\mathcal{H}_{U}+\mathcal{H}_{L})\left(1-\frac{m^2_{\ell}}{2q^2}\right)
-\frac{3m^2_{\ell}}{2q^2}\mathcal{H}_{S}}{\mathcal{H}_\mathrm{total}},
\\
P_\mathrm{T}^\ell(q^2)
&=-\frac{3\pi m_\ell}{8\sqrt{q^2}}\frac{\mathcal{H}_{P}+2\mathcal{H}_{SL}}
{\mathcal{H}_\mathrm{total}},
\\
F_\mathrm{L}^\ell(q^2)
&=\frac{\mathcal{H}_{L}\left(1+\frac{m^2_{\ell}}{2q^2}\right)+\frac{3m^2_{\ell}}{2q^2}\mathcal{H}_{S}}
{\mathcal{H}_\mathrm{total}}.
\label{Eq:FL}
\end{align}
In which, the total helicity amplitude is
\begin{align}
\mathcal{H} _{\mathrm{total}}=(\mathcal{H} _U+\mathcal{H} _L)\left( 1+\frac{m_{\ell}^{2}}{2q^2} \right) +\frac{3m_{\ell}^{2}}{2q^2}\mathcal{H} _S.
\end{align}
After integrating over $\cos\theta$, the differential decay distribution changed with the squared momentum transfer $q^2$ can be written as
\begin{align}
\frac{d\Gamma (D^+\to \omega\ell ^+\nu _{\ell})}{dq^2}=\frac{G_{F}^{2}}{(2\pi )^3}|V_{cd}|^2\frac{\lambda ^{1/2}(q^2-m_{\ell}^{2})^2}{24M_{D^+}^{3}q^2}\mathcal{H} _{\mathrm{total}}.
\label{Eq:dgamma}
\end{align}
Here, due to chiral suppression, the leptonic mass $m_{\ell}$ can be neglected. Then, the differential decay width can be decomposed into longitudinal polarization state and transverse polarization state
\begin{align}
\frac{d\Gamma_{\rm{L}} (D^+\to \omega\ell ^+\nu _{\ell})}{dq^2}&=\frac{G_{F}^{2}}{(2\pi )^3}|V_{cd}|^2\frac{\lambda ^{1/2}q^2}{24M_{D^+}^{3}}\mathcal{H} _L,\nonumber\\
\frac{d\Gamma_{\rm{T}} (D^+\to \omega\ell ^+\nu _{\ell})}{dq^2}&=\frac{G_{F}^{2}}{(2\pi )^3}|V_{cd}|^2\frac{\lambda ^{1/2}q^2}{24M_{D^+}^{3}}\mathcal{H} _U.
\label{Eq:dLT}
\end{align}
The branching fraction can be obtained by
\begin{align}
\mathcal{B}(D^+\to \omega\ell\nu_{\ell}) = \frac{\tau_{D^+}}{c^2_{\omega}}\int^{q^2_{\rm{max}}}_{m_{\ell}^2} \frac{d\Gamma (D^+\to \omega\ell ^+\nu _{\ell})}{dq^2},
\label{Eq:Br}
\end{align}
where $q^2_{\rm{max}}=(m_{D^+}-m_{\omega})^2$ and $c_{\omega}=\sqrt{2}$ is isospin factor from flavor wave function of the $\omega$-meson, $(u\bar{u}+d\bar{d})/\sqrt{2}$. Furthermore, one can usually decompose them into sequential decay processes via factorization relations when calculating multibody decay processes~\cite{Cheng:2020iwk,Wu:2025dio,Cheng:2022vbw}. This method is valid only in the narrow width limit. Given this, we can estimate the branching fraction for five body decay $\mathcal{B}(D^+\to \omega(\to \pi^+\pi^-\pi^0)\ell^+\nu_{\ell})$ by following relation
\begin{align}
\mathcal{B}(D^+\to &\omega(\to \pi^+\pi^-\pi^0)\ell^+\nu_{\ell})\nonumber\\
&=\mathcal{B}(D^+\to \omega\ell^+\nu_{\ell})\times \mathcal{B}(\omega\to \pi^+\pi^-\pi^0).
\label{Eq:Br pipipi}
\end{align}

Next step, for the TFFs originating from the hadronic matrix element, it is difficult to calculate by pure perturbative methods due to the presence of color confinement and large strong coupling constant in the low energy region. As mentioned in Section~\ref{Sec:I}, we employ the LCSR approach for the calculation. As an effective combination of SVZ sum rules and hard exclusive process, we can first start from the vacuum to meson correlation function,
\begin{align}
\Pi _{\mu}(p,q)&=i\int d^4xe^{iq\cdot x}\langle \omega (p,\lambda )|\mathrm{T}\{ \bar{q}_1(x)\gamma _{\mu}(1-\gamma _5)c(x),
\nonumber\\
&\times j_{D^+}^{\dagger}(0)\} |0\rangle.
\label{Eq:correlation function}
\end{align}
Here, we adopt the right-handed chiral current $j_{D^+}^{\dagger}(x)=i\bar{c}(x)(1+\gamma_5)q_2(x)$. According to the basic steps of LCSR method, the correlation function can insert a complete intermediate state with the same quantum number as the current operator $i\bar{c}(x)(1+\gamma_5)q_2(x)$ into the hadron current in the timelike $q^2$ region. After separating the pole term of the lowest $D^+$-meson and replacing the contributions from higher resonances and continuum states with dispersion relation, the hadronic representation of correlation function can be obtained. Secondly, based on QCD theory, the correlation function can be carried out OPE near the light-cone $x^2\rightsquigarrow 0$ in the space-like $q^2$ region. Finally, with the help of quark hadron duality and Borel transformation, the analytic expression of TFFs can be obtained. After our verification, this result is similar to our previous work on $B\to \rho$~\cite{Fu:2014pba}, while the hadronic representation gives different hadronic parameters accordingly. Here, we do not provide the specific expressions and our focus will be on discussing twist-2 LCDA $\phi^{\perp}_{2;\omega}(x,\mu)$.

The integral relation between twist-2 $\phi^{\perp}_{2;\omega}(x,\mu)$ and the WF of valence Fock state can be defined as
\begin{align}
\phi _{2;\omega}^{\bot}(x,\mu)=\frac{2\sqrt{3}}{\widetilde{f}_{\omega}^{\bot}}\int_{|\mathbf{k}_{\bot}|^2\le \mu _{0}^{2}}{\frac{d\mathbf{k}_{\bot}}{16\pi ^3}\Psi _{2;\omega}^{\bot}(x,\mathbf{k}_{\bot}),}
\label{Eq:twist2 Integrate}
\end{align}
with $\widetilde{f}_{\omega}^{\bot}=f_{\omega}^{\bot}/\sqrt{3}$.  Based on the BHL description, the LCWF can be expressed as
\begin{align}
\Psi _{2;\omega}^{\bot}(x,\mathbf{k}_{\bot})=\sum_{h_1h_2}{\chi _{\omega}^{h_1h_2}(x,\mathbf{k}_{\bot})\Psi_{2;\omega}^{R}(x,\mathbf{k}_{\bot})}
\end{align}
In which $\lambda_1$ and $\lambda_2$ are the helicity of $q_1$ and $\bar{q}_2$ in spin WF $\chi _{\omega}^{h_1h_2}(x,\mathbf{k}_{\bot})$, respectively. By using the Wigner-Melosh rotation~\cite{Cao:1997hw,Cao:1997sx,Melosh:1974cu}, the different helicities of $\chi _{\omega}^{h_1h_2}(x,\mathbf{k}_{\bot})$ can be obtained
\begin{align}
\sum_{h_1h_2}\chi _{\omega}^{h_1h_2}(x,\mathbf{k}_{\bot})=\frac{\hat{m}_q}{\sqrt{\mathbf{k}_{\bot}^2+\hat{m}_q^2}}
\end{align}
where $\hat{m}_q=300~\rm{GeV}$ is  constituent quark mass. For spatial WF $\Psi_{2;\omega}^{R}(x,\mathbf{k}_{\bot})$, since the LCWF must satisfy an infinite set of coupled integral equations, it is difficult to obtain its exact solution. To address this situation, the BHL assumes that the energy in the conventional coordinate system (C.M.) is equal to the off-shell energy in the infinite momentum frame (L.C.)~\cite{Guo:1991eb,Huang:1994dy,Huang:1980yra}, $.ie.,$
\begin{align}
\varepsilon =
\begin{cases}
M^2 - \left( \sum_{i=1}^{n} q_i^0 \right)^2, & \sum_{i=1}^{n} \mathbf{q}_i = 0 \quad [{\rm C.M.}] \\
\nonumber\\
M^2 - \sum_{i=1}^{n} \left( \dfrac{\mathbf{k}_{\perp i}^2 + m_i^2} {x_i} \right), & \begin{cases}\sum_{i=1}^{n} \mathbf{k}_{\perp i} = 0 \\
\sum_{i=1}^{n} x_i = 1 \quad  \end{cases} \quad [{\rm L.C.}]
\end{cases}
\end{align}
The solution of the two-body Bethe-Salpeter bound state wave function in the case of weakly bound states shows that the solution of the equal-time wave function in the C.M. frame  is a function of energy $\varepsilon$. Consequently, for a two-particle bound state, the LCWF and the equal-time WF have the following relationship~\cite{Wu:2011gf,Sun:2010oyk}
\begin{align}
\psi_{\rm L.C.}\left( \frac{\mathbf{k}_{\bot}^2 + m_q^2}{4x\bar{x}}  \right) \leftrightarrow \psi_{\rm C.M.}\left( \mathbf{q}^2 \right)
\end{align}
with $\bar{x}=(1-x)$. Then, the  spatial WF of LCHO model can be obtained
\begin{align}
\Psi^{R}_{2;\omega}(x,\mathbf{k}_{\bot})&= A_{2;\omega}^{\perp}
\exp \left[ -b_{2;\omega}^{\perp 2} \frac{\mathbf{k}_{\bot}^{2}+\hat{m}_q^2}{x\bar{x}}\right],
\end{align}
where $ A_{2;\omega}^{\perp}$ is normalization constant and $b_{2;\omega}^{\perp}$ is the harmonic parameter. In addition, when we take into account the existence of quark spins, Lorentz boost will introduce additional longitudinal corrections to LCDA. Therefore, it is necessary to introduce an additional longitudinal correction function $\varphi(x)$ into LCDA to deal with it, which can be expanded in Gegenbauer polynomials. The specific form is
\begin{align}
\varphi(x)=1+B^{\perp}_{2;\omega}C^{3/2}_2 (2x-1).
\end{align}
In which $C^{3/2}_2 (2x-1)$ is Gegenbauer polynomial and $B^{\perp}_{2;\omega}$ dominates the longitudinal distribution. Finally, by  using Eq.~\eqref{Eq:twist2 Integrate} to integrate over transverse momentum, the LCHO model of twist-2 LCDA $\phi_{2\omega}^{\bot}(x,\mu)$ can be determined, and which reads
\begin{align}
&\phi _{2;\omega}^{\bot}(x,\mu )=\frac{A_{2;\omega}^{\bot}\sqrt{3x\bar{x}}\hat{m}_q}{8\pi ^{3/2}\widetilde{f}_{\omega}^{\bot}b_{2;\omega}^{\bot}}[1+B_{2;\omega}^{\bot}C_{2}^{3/2}(\xi )]\nonumber\\
&\times \left[ \mathrm{Erf}\left( b_{2;\omega}^{\bot}\sqrt{\frac{\mu ^{2}+\hat{m}_{q}^{2}}{x\bar{x}}} \right) -\mathrm{Erf}\left( b_{2;\omega}^{\bot}\sqrt{\frac{\hat{m}_{q}^{2}}{x\bar{x}}} \right) \right].
\end{align}
where $\rm{Erf}$$(x) =2\int_0^x{e^{-t^2}dx/\sqrt{\pi}}$ is the error function. From this form, two advantages can be reflected: (i) It has a form close to the asymptotic behavior $\phi _{2;\omega}^{\bot}(x,\mu \rightarrow \infty)=6x\bar{x}$; (ii) It has an exponentially suppressed endpoint behavior, which eliminates endpoint singularity well in convolution. The determination of the remaining three model parameters $A_{2;\omega}^{\bot}, b_{2;\omega}^{\bot}$ and $B_{2;\omega}^{\bot}$ requires the application of the three conditions: the condition for the strict normalization of twist-2 LCDA $\phi _{2;\omega}^{\bot}(x,\mu )$, the average value of the squared transverse momentum $\langle \mathbf{k}_{\bot}^2 \rangle_{2;\omega}$
\begin{align}
\langle \mathbf{k}_{\bot}^2 \rangle _{2;\omega}=\frac{\int{dxd^2\mathbf{k}_{\bot}|\mathbf{k}_{\bot}|^2|\Psi _{2;\omega}^{\bot} (x,\mathbf{k}_{\bot})|^2}}{\int{dxd^2\mathbf{k}_{\bot}|\Psi _{2;\omega}^{\bot}(x,\mathbf{k}_{\bot})|^2}}.
\end{align}
where we use $\langle \mathbf{k}_{\bot}^2 \rangle _{2;\omega}^{1/2}=0.37~{\rm GeV}$ from Refs.~\cite{Wu:2013lga,Guo:1991eb}, and the connection between the first Gegenbauer moments and twist-2 $\phi _{2;\omega}^{\bot}(x,\mu )$
\begin{align}
a_{2;\omega}^{\bot} (\mu)=\frac{\int_0^1{dx\phi _{2;\omega}^{\bot} (x,\mu)C_{2}^{3/2}(2x-1)}}{\int_0^1{6x\bar{x}[C_{2}^{3/2}(2x-1)]^2}},
\label{Eq:a2}
\end{align}
for which, we take the value of $a_{2;\omega}^{\bot} (\mu_0)=0.14\pm0.12$ from DL~\cite{Dimou:2012un}.

\section{Numerical Analysis}\label{Sec:III}
\begin{figure}[t]
\begin{center}
\includegraphics[width=0.435\textwidth]{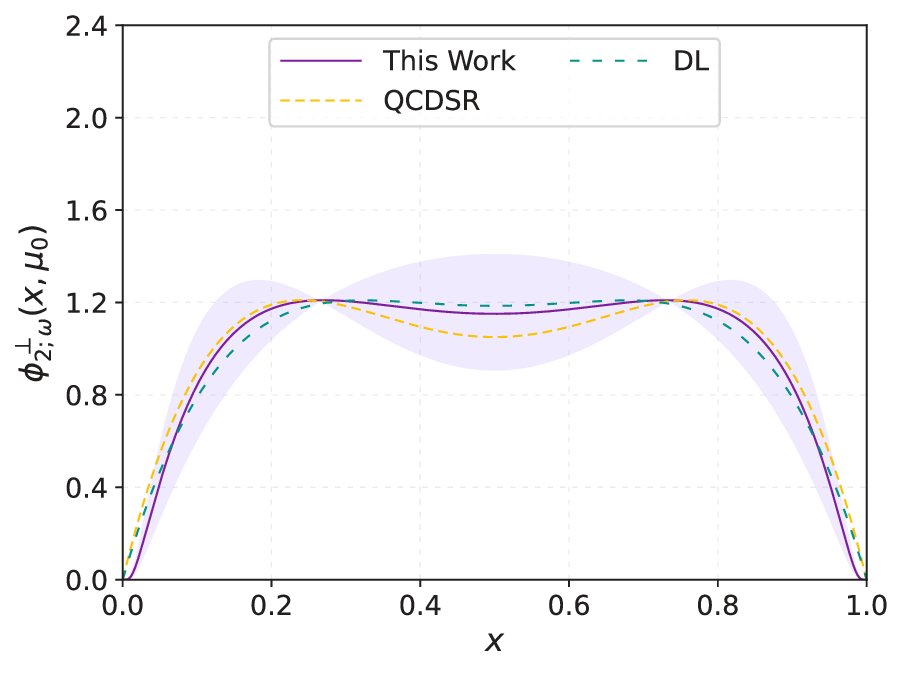}
\caption{The comparison of $\phi _{2;\omega}^{\bot} (x,\mu_0)$ with QCDSR~\cite{Ball:1998sk} and DL~\cite{Dimou:2012un} at $\mu_0=1~\rm{GeV}$, where the shaded band represents the uncertainty.}
\label{Fig:DA}
\end{center}
\end{figure}

In order to do numerical calculation, the basic input parameters are taken from PDG~\cite{ParticleDataGroup:2024cfk}: the mass of $D^+$ and $\omega$-mesons are $m_{D^+}=1869.66\pm0.05~\rm{MeV}$ and $m_{\omega}=782.66\pm0.13~\rm{MeV}$, the charm quark mass is $m_c(\bar{m}_c)=1.2730\pm0.0046~\rm{GeV}$, the $D^+$-meson decay constant is $f_{D^+}=212.0\pm0.7~\rm{MeV}$. And the $m_{\omega}$-meson decay constant is $f_{\omega}^{\bot}=145\pm10~\rm{MeV}$~\cite{Ball:2004rg}. In this process, the typical process energy scale $\mu_{k}=\sqrt{m^2_{D^+}-m^2_c}\approx 1.4~\rm{GeV}$. Before calculating TFFs, since the twist-2 LCDA $\phi _{2;\omega}^{\bot}(x,\mu )$ serves as one of the crucial input parameters, we first need to determine its precise behavior. With the help of QCD evolution~\cite{Ball:2003sc}, the Gegenbauer moments $a_{2;\omega}^{\bot} (\mu)$ can be evolved from the initial energy scale $\mu_0$ to $\mu_k$. Then, the three model parameters of $\phi _{2;\omega}^{\bot}(x,\mu )$ can be determined and listed in Table~\ref{table:modelparameter}, where include the variations induced by the uncertainty of $a_{2;\omega}^{\bot} (\mu_0)$. Meanwhile, in Fig.~\ref{Fig:DA}, we present a comparison between our twist-2 $\phi _{2;\omega}^{\bot}(x,\mu_0 )$ behavior and the results from QCDSR~\cite{Ball:1998sk} and DL~\cite{Dimou:2012un}. Notably, both of their DA models utilize the conventional Gegenbauer polynomial expansion. As seen in Fig.~\ref{Fig:DA}, the central result of our LCHO model central result agrees well with other QCDSR and DL. Moreover, although the contribution of high twists will be involved in our TFFs, this part will be highly suppressed in sum rule method. The corresponding expressions and input parameters can be found in Refs~\cite{Fu:2014pba,Wu:2006rd,Fu:2020vqd}.
\begin{table}
\begin{center}
\renewcommand{\arraystretch}{1.5}
\caption{The results of $\omega$-meson twist-2 LCDA $\phi _{2;\omega}^{\bot} (x,\mu)$ parameters $A_{2;\omega}^{\bot} ~({\rm GeV}^{-1})$, $b_{2,\omega}^{\bot} ~({\rm GeV}^{-1})$ and $B_{2;\omega}^{\bot}$, which is corresponded to the upper limit, central value, and lower limit of $a_{2;\omega}^{\bot} (\mu)$ at $\mu_0$ and $\mu_k$, respectively.}
\label{table:modelparameter}
\begin{tabular}{l c@{\hspace{2em}} c@{\hspace{2em}} c}
\hline
        &$A_{2;\omega}^{\bot} $      &$b_{2;\omega}^{\bot} $    &$B_{2;\omega}^{\bot} $  \\
\hline
                     &20.618                   &~~0.563                    &0.216      \\           \cline{2-4}
$\mu_0$ ~~~            &22.619                   &~~0.596                    &0.112   \\            \cline{2-4}
                     &24.546                   &-0.628                    &0.002    \\             \cline{2-4}
\hline
                     &19.123                  &-0.564                    &0.213  \\                \cline{2-4}
$\mu_k$  ~~~            &20.968                   &-0.595                    &0.117   \\          \cline{2-4}
                     &22.770                   &-0.624                    &0.017    \\             \cline{2-4}
\hline
\end{tabular}
\end{center}
\end{table}
\begin{figure*}[t]
\begin{center}
\includegraphics[width=0.4\textwidth]{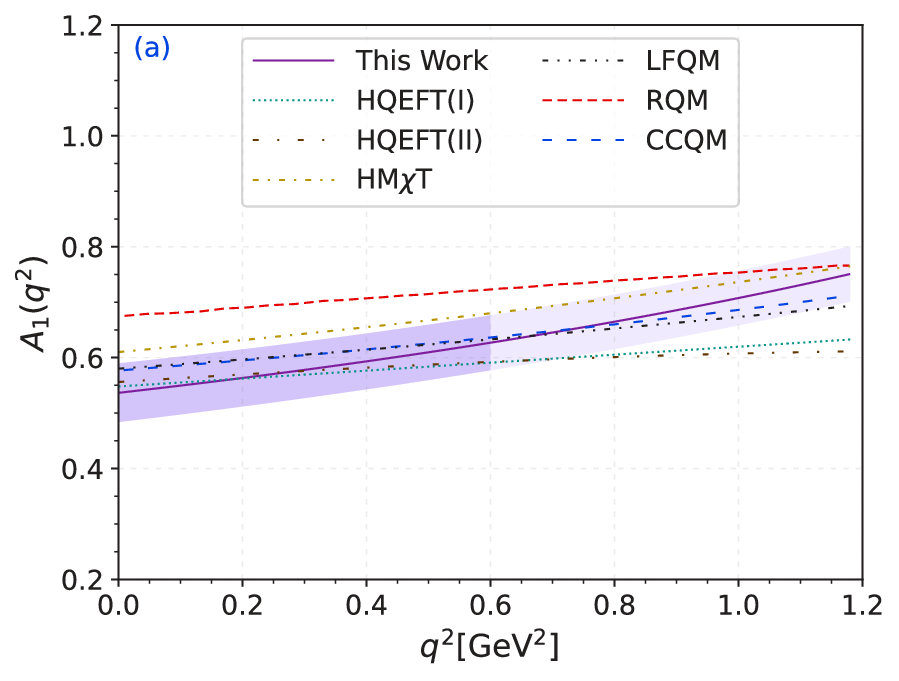}
\includegraphics[width=0.4\textwidth]{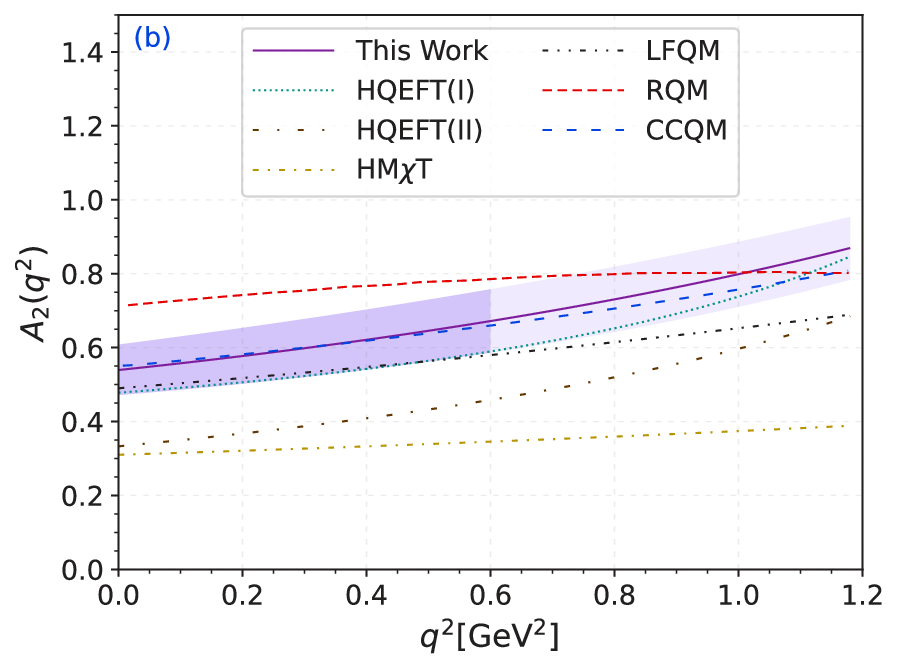}
\includegraphics[width=0.4\textwidth]{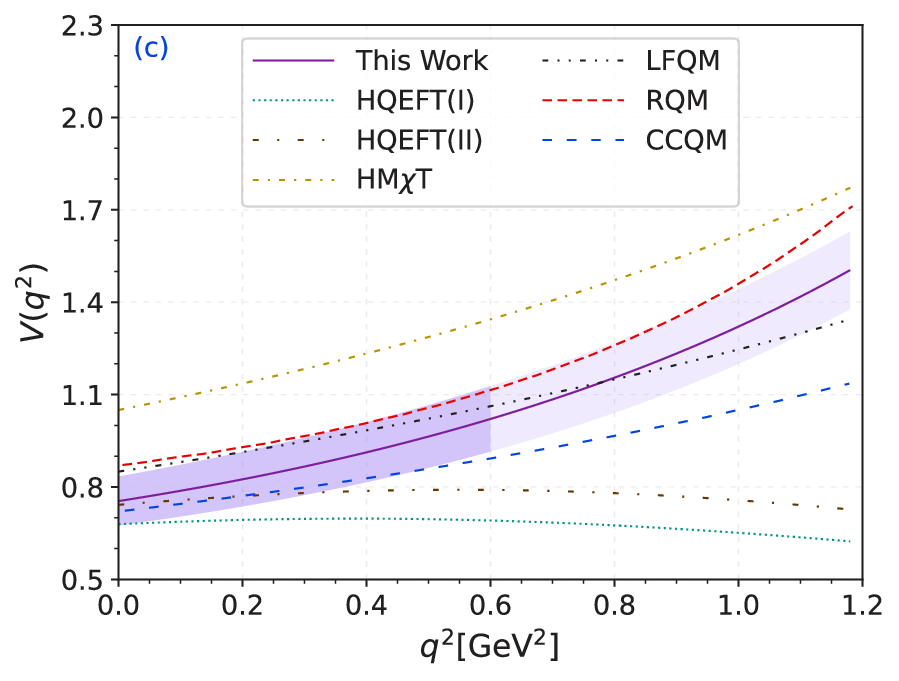}
\includegraphics[width=0.4\textwidth]{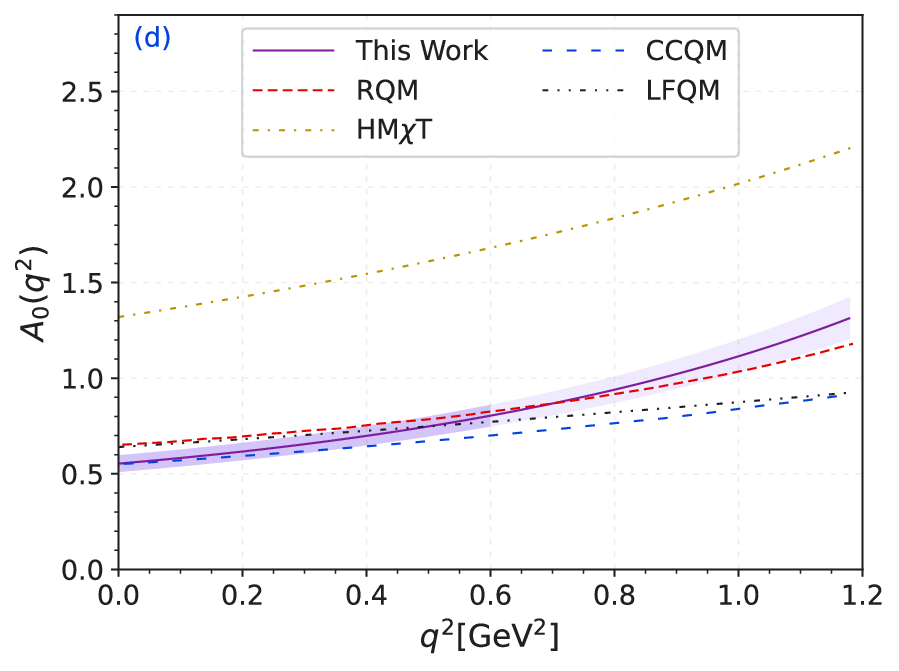}\\
\end{center}
\caption{The behavior of TFFs (a) $A_1(q^2)$, (b) $A_2(q^2)$, (c) $V(q^2)$, and (d) $A_0(q^2)$. For comparison, the predictions from HQEFT~\cite{Wu:2006rd}, $\rm{HM}\chi\rm{T}$~\cite{Fajfer:2005ug}, LFQM~\cite{Verma:2011yw}, RQM~\cite{Faustov:2019mqr} and CCQM~\cite{Ivanov:2019nqd} are also presented.}
\label{Fig:TFFscomparison}
\end{figure*}
\begin{table}[h]
\renewcommand{\arraystretch}{1.5}
\begin{center}
\caption{Comparison of TFFs $A_1(0), A_2(0)$ and $V(0)$ at large recoil point}
\label{table:lparge point}
\begin{tabular}{l c c c c}
\hline
                          &$A_1(0)$ & $~A_2(0)$ & $~V(0)$   &$A(0)$\\
\hline
This work                 &$0.537^{+0.053}_{-0.053}$ & $0.540^{+0.068}_{-0.068}$ & $0.754^{+0.079}_{-0.079}$  & $~~~0.553^{+0.044}_{-0.043}$\\
HQEFT(I)~\cite{Wu:2006rd}   &$0.548^{+0.029}_{-0.027}$ & $0.478^{+0.034}_{-0.029}$& $0.679^{+0.030}_{-0.023}$ &$-0.478^{+0.029}_{-0.034}$\\
HQEFT(II)~\cite{Wu:2006rd}   &$0.556^{+0.033}_{-0.028}$ & $0.333^{+0.026}_{-0.030}$& $0.742^{+0.041}_{-0.034}$  &$-0.657^{+0.053}_{-0.065}$\\
HM$\chi$T~\cite{Fajfer:2005ug} &$~0.61$ & $~0.31$ & $1.05$   &$1.32$\\
LFQM~\cite{Verma:2011yw}     &$0.58$ & $0.49$ & $0.85$   &$0.64$\\
RQM~\cite{Faustov:2019mqr}   &$0.674$ & $0.713$ & $0.871$ &$0.647$\\
CCQM~\cite{Ivanov:2019nqd}    &$0.58$ & $0.55$ & $0.72$ &$\cdot\cdot\cdot$\\
 \hline
\end{tabular}
\end{center}
\end{table}

There are two other important parameters: continuum threshold $s_0$ and Borel parameter $M^2$. Based on the basic criteria of the sum rules approach~\cite{Hu:2021zmy}, we take $s^{A_1}_0= 5.7\pm0.1~{\rm GeV}^2, s^{A_2}_0= 5.5\pm0.1~{\rm GeV}^2, s^{V}_0= 5.0\pm0.1~{\rm GeV}^2, s^{A_0}_0= 6.8\pm0.1~{\rm GeV}^2, M^2_{A_1}=6.7\pm0.1~{\rm GeV}^2, M^2_{A_2}=4.0\pm0.1~{\rm GeV}^2$, $M^2_V=6.0\pm0.1~{\rm GeV}^2$ and $M^2_{A_0}=6.5\pm0.1~{\rm GeV}^2$. With above parameters, we can calculate the TFFs. The predictions for TFFs $A_1(0), A_2(0)$, $V(0)$ and $A_0(0)$ at the large recoil point are presented in Table~\ref{table:lparge point}, where also includes the results from HQEFT~\cite{Wu:2006rd}, HM$\chi$T~\cite{Fajfer:2005ug}, LFQM~\cite{Verma:2011yw}, RQM~\cite{Faustov:2019mqr} and CCQM~\cite{Ivanov:2019nqd}. In which, the HQEFT~\cite{Wu:2006rd} provides two results. Specifically, HQEFT(I) only considers the leading-twist meson DAs, while HQEFT(II) takes into account the meson DAs up to twist-4. As shown in Table~\ref{table:lparge point}, the result of $A_1(0)$ is consistent with predictions from other theoretical groups. The central values of $A_2(0)$ and $V(0)$ show minor deviations from other theoretical predictions, and are in good agreement within uncertainties. However, a significant discrepancy is observed for $A(0)$. In fact, among these four TFFs, we are primarily concerned with $A_1(q^2)$, $A_2(q^2)$, and $V(q^2)$. For $A_0(q^2)$, when we calculate the physical observables, it is always accompanied by a coefficient $m_{\ell}^2/q^2$, which makes its contribution strongly suppressed. In addition, nearly all helicity amplitudes contain the TFF $A_1(q^2)$. Then the large recoil region, we can define two important TFF ratios, $r_V=V(0)/A_1(0)$ and $r_2=A_2(0)/A_1(0)$. These ratios can be obtained without making any assumptions about the total decay width or CKM matrix elements, which makes them of great interest to experiments as well. The following is our prediction
\begin{align}
&r_V=1.40^{+0.21}_{-0.19}, &&r_2=1.01^{+0.17}_{-0.16}.
\end{align}
Our predictions agree well with BESIII Collaboration~\cite{BESIII:2015kin}, and their results have been mentioned earlier in introduction.

Since the LCSR method is only applicable in the low and intermediate $q^2$ region, $i.e., q^2\in [0, 0.6]$, while decay widths and branching fraction require the global behavior of TFFs, we subsequently employ the simplified series expansion (SSE) method to extend the TFFs to the whole kinematical region, which is defined as~\cite{Bourrely:2008za}
\begin{align}
F_{i}(q^2)=\frac{1}{1-q^2/m^2_{R,i}}\sum_{k=0,1,2}\beta_{k,i} z^k(q^2,t_0).
\label{Eq:SSE}
\end{align}
\begin{table}[h]
\renewcommand{\arraystretch}{1.5}
\begin{center}
\caption{The central value of fitted parameters $m_{R,i}$, $\beta_{k,i}$ and quality of extrapolation $\Delta$ for $D^+\to \omega$ TFFs $A_1(q^2), A_2(q^2)$, $V(q^2)$, and $A_0(q^2)$.}
\label{table:fitParameter}
\begin{tabular}{l@{\hspace{2em}} c@{\hspace{2em}} c @{\hspace{2em}}c @{\hspace{2em}}c}
\hline
   &$ A_1(q^2)$  & $~A_2(q^2)$  & $~V(q^2)$  &$A_0(q^2)$\\
\hline
$m_{R,i}$   &~$2.422$     &~2.422   &~2.006  &~$2.422$\\
$\beta_{0,i}$   &~0.537     &~0.540   &~0.754  &~$0.553$\\
$\beta_{1,i}$   &-0.991     &-2.367  &-5.204   &-4.963\\
$\beta_{2,i}$   &~8.401    &~21.951   &~92.184  &~$115.247$\\
$\Delta$    &$~0.119\%$  &$~0.001\%$  &$~0.024\%$  &~$0.053\%$\\
\hline
\end{tabular}
\end{center}
\end{table}
$F_{i}(q^2)$ represents four TFFs $A_{1}(q^2), A_{2}(q^2)$, $V(q^2)$ and $A_{0}(q^2)$. $\beta_{k,i}$ are real coefficients, while the function $z^k(q^2,t_0)=(\sqrt{t_+-q^2}-\sqrt{t_+-t_0})/(\sqrt{t_+-q^2}+\sqrt{t_+-t_0})$ with $t_{\pm}=(m_{D^+}\pm m_\omega )^2$ and $t_0=t_+(1-\sqrt{1-t_-/t_+})$. This is a systematic and model independent parameterization approach for semileptonic TFFs. It can easily translate the near threshold behavior of TFFs into useful constraints on the expansion coefficients, while also ensuring the analytic structure of TFFs. We need to assess the reasonableness of the fitted curve by a quality of extrapolation, $\Delta =\sum_t{|F_i(t)-F_i^{\rm fit}(t)|}/{\sum_t{|F_i(t)|}}~\times 100$. The real coefficients $\beta_{k,i}$ can be naturally determined by imposing the requirement that $\Delta< 1\%$. Meanwhile, according to different quantum numbers $J^P$, the masses of resonances are $m_{R,i}=2.006~\rm{GeV}$ for TFFs $V(q^2)$ with $J^P=1^-$ and $m_{R,i}=2.422~\rm{GeV}$ for TFFs $A_{1,2,0}(q^2)$ with $J^P=1^+$, respectively~\cite{ParticleDataGroup:2024cfk}. Then, the fitted parameters can be determined and presented in Table~\ref{table:fitParameter}, where the all the LCSR parameters set to be their central values. We can observe that $\Delta$ is significantly less than $1\%$, which indicates that the extrapolation effect is excellent. With the above definitions, the behavior of the TFFs can be determined in the whole $q^2$ region, which is shown in Fig.~\ref{Fig:TFFscomparison}. The predictions from HQEFT~\cite{Wu:2006rd}, $\rm{HM}\chi\rm{T}$~\cite{Fajfer:2005ug}, LFQM~\cite{Verma:2011yw}, RQM~\cite{Faustov:2019mqr} and CCQM~\cite{Ivanov:2019nqd} are also included. Due to the different approaches, the extrapolation trends also show variations. For $A_1(q^2)$, in the low and intermediate $q^2$ region, it well encompasses the majority of theoretical results. Within errors, it shows an agreement with CCQM and HQEFT(II). For $A_2(q^2)$, the overall behavior agrees with CCQM and HQEFT(I). As for $V(q^2)$ and $A_0(q^2)$, our predictions show good agreement with RQM.

\begin{figure*}[t]
\begin{center}
\includegraphics[width=0.4\textwidth]{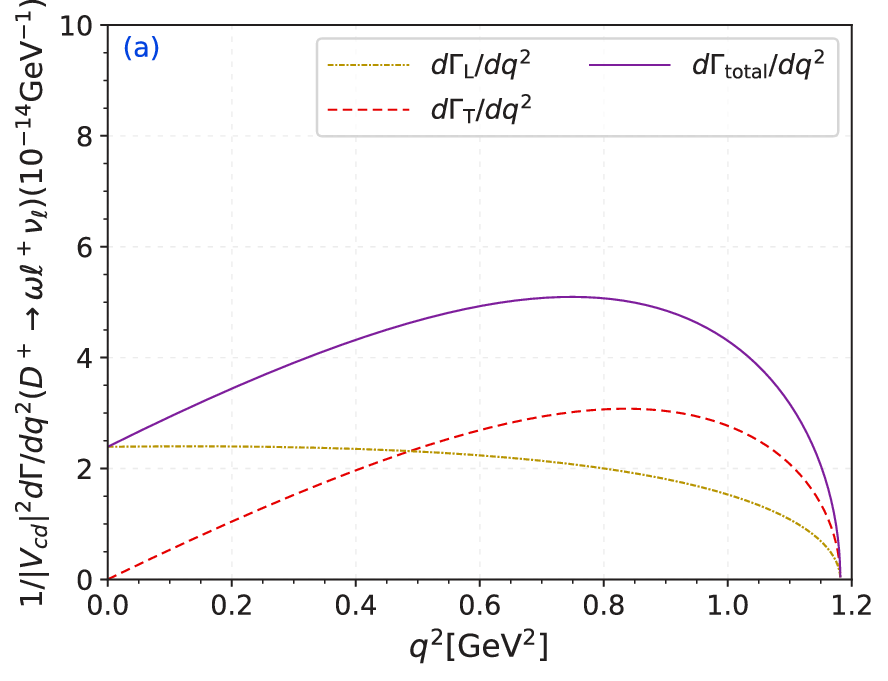}
\includegraphics[width=0.4\textwidth]{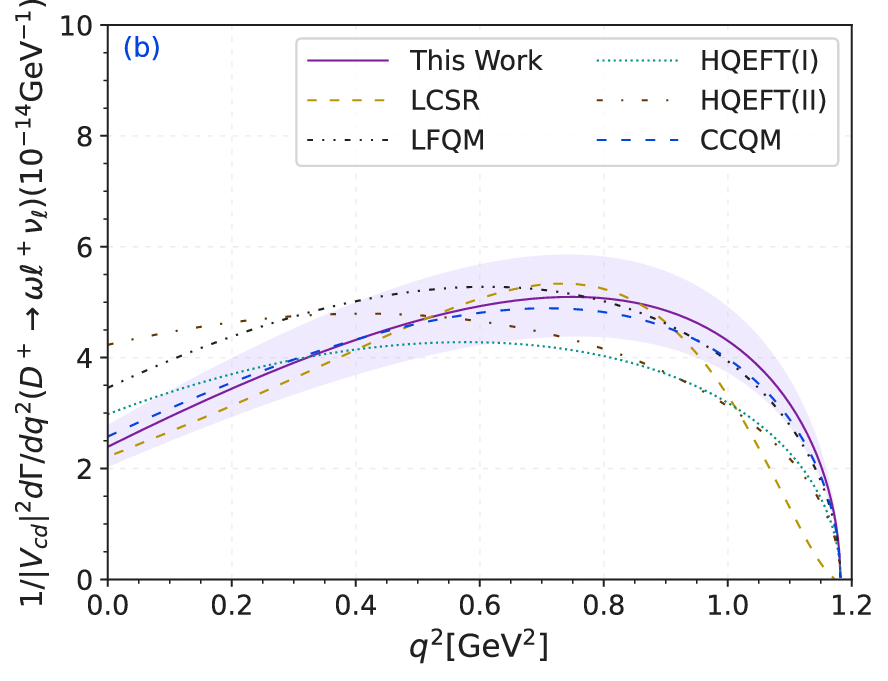}
\end{center}
\caption{The differential decay width $1/|V_{cd}|^2d\Gamma(D^+\to \omega\ell^+\nu_{\ell})/dq^2$ as a function of $q^2$, where (a) denotes the central values of longitudinal, transverse and total CKM-independent differential decay width, (b) denotes the comparison of various experimental and theoretical results for total CKM-independent differential decay width.}
\label{Fig:dGamma}
\end{figure*}
Next step, we can calculate the CKM-independent differential decay width $1/|V_{cd}|^2d\Gamma(D^+\to \omega\ell^+\nu_{\ell})/dq^2$. First, the central values of longitudinal, transverse and total CKM-independent differential decay width are presented in Fig.~\ref{Fig:dGamma} ({\color{blue}{a}}). For comparison, the total CKM-independent differential decay width of theoretical predictions from LCSR~\cite{Fu:2020vqd}, LFQM~\cite{Cheng:2017pcq}, HQEFT~\cite{Wu:2006rd}, and CCQM~\cite{Soni:2018adu} are all included in Fig.~\ref{Fig:dGamma} ({\color{blue}{b}}). Most theoretical groups have not provided corresponding predictions for this observable. Here, we give their central predictions by fitting their TFFs. In the region near $q^2=0$, our central result shows significant difference from those of HQEFT(II) and LFQM, which can be attributed to the variation in $A_2(0)$.  Overall, the results of CCQM are show good agreement with our prediction across the whole $q^2$ region. After integrating over $q^2$ in entire physical region, the corresponding decay widths can be obtained (in GeV)
\begin{align}
\Gamma_{\rm{L}}(D^+\to\omega\ell^+\nu_{\ell})=(2.395^{+0.633}_{-0.567})\times 10^{-14},\nonumber\\
\Gamma_{\rm{T}}(D^+\to\omega\ell^+\nu_{\ell})=(2.427^{+0.339}_{-0.315})\times 10^{-14},\nonumber\\
\Gamma_{\rm{total}}(D^+\to\omega\ell^+\nu_{\ell})=(4.822^{+0.951}_{-0.861})\times 10^{-14}.
\end{align}
This leads to $\Gamma_{\rm{L}}/\Gamma_{\rm{T}}=0.987^{+0.107}_{-0.121}$. Then, by using the lifetime $\tau_{D^+}=(1.033\pm0.005)~\rm{ps}$ and CKM matrix element $|V_{cd}|=0.221\pm0.004$ from PDG~\cite{ParticleDataGroup:2024cfk}, and integrating over $q^2$ in $m^2_{\ell} \leq q^2 \leq (m_{D^+}-m_{\omega})$, the branching fraction $\mathcal{B}(D^+\to \omega\ell^+\nu_{\ell})$ with $\ell=(e,\mu)$ can be determined. Table~\ref{table:Br} presents a comparison between our results and those from other theoretical and experimental studies. Our results are in good agreement with those from other groups at the order of $10^{-3}$. For example, the prediction of $\mathcal{B}(D^+\to \omega e^+\nu_e)$ shows good agreement with those from CLEO'11~\cite{CLEO:2011ab} and CCQM~\cite{Soni:2018adu}, while $\mathcal{B}(D^+\to \omega \mu^+\nu_{\mu})$ result shows better consistency, which is consistent with the results of PDG~\cite{ParticleDataGroup:2024cfk}, BESIII'20~\cite{Ablikim:2020tmg}, CCQM~\cite{Soni:2018adu} and LCSR~\cite{Fu:2020vqd}. It can also be clearly observed here that the world average results provided by PDG show that the branching fraction of the electron channel is smaller than that of muon channel, while theoretical predictions strictly adhere to the conclusion that the branching fraction of electron channel is larger than that of muon channel.
\begin{table}[h]
\renewcommand{\arraystretch}{1.5}
\begin{center}
\caption{Comparison of various experimental and theoretical results for the $D^+\to \omega\ell^+\nu_{\ell}$ branching fraction within uncertainties (in unit $10^{-3}$)}
\label{table:Br}
\begin{tabular}{l@{\hspace{2em}}llll}
\hline
   &$\mathcal{B}(D^+\to \omega e^+\nu_e)$ &$\mathcal{B}(D^+\to \omega \mu^+\nu_{\mu})$\\
\hline
This work   &$1.848^{+0.365}_{-0.330}$~~~~~&$1.782^{+0.334}_{-0.303}$\\
PDG~\cite{ParticleDataGroup:2024cfk}   &$1.69\pm0.11$~~~~~&$1.77\pm0.21$\\
BESIII'15~\cite{BESIII:2015kin}&$1.63\pm0.11\pm0.08$~~~~&$--$\\
BESIII'20~\cite{Ablikim:2020tmg}&$--$~~~~&$1.77\pm0.18\pm0.11$\\
CLEO'05~\cite{CLEO:2005rxg}&$1.6^{+0.7}_{-0.6}\pm0.1$~~~~&$--$\\
CLEO'11~\cite{CLEO:2011ab}&$1.82^{+0.18}_{-0.07}\pm0.01$~~~~&$--$\\
HQEFT(I)~\cite{Wu:2006rd}   &$1.72^{+0.15}_{-0.14}$~~~~~&$1.65^{+0.14}_{-0.13}$\\
HQEFT(II)~\cite{Wu:2006rd}   &$1.93^{+0.20}_{-0.14}$~~~~~&$1.85^{+0.19}_{-0.13}$\\
$\rm{HM}\chi\rm{T}$~\cite{Fajfer:2005ug}&$2.5$~~~~~&$--$\\
LFQM~\cite{Cheng:2017pcq}&$2.1\pm0.2$~~~~&$2.0\pm0.2$\\
$\chi$UA~\cite{Sekihara:2015iha}&$2.46$~~~~&$2.29$\\
CCQM~\cite{Soni:2018adu}&$1.85$~~~~&$1.78$\\
LCSR~\cite{Fu:2020vqd}&$1.74^{+0.482}_{-0.399}$~~~~&$1.728^{+0.479}_{-0.397}$\\
RQM~\cite{Faustov:2019mqr}&$2.17$~~~~&$2.08$\\
 \hline
\end{tabular}
\end{center}
\end{table}
Additionally, we predicted the branching fraction for $D^+\to \omega(\to \pi^+\pi^-\pi^0)\ell^+\nu_{\ell}$ by Eq.~\eqref{Eq:Br pipipi},
\begin{align}
\mathcal{B}(D^+\to \omega(\to \pi^+\pi^-\pi^0)e^+\nu_{e})=1.648^{+0.341}_{-0.305}\times 10^{-3},\nonumber\\
\mathcal{B}(D^+\to \omega(\to \pi^+\pi^-\pi^0)\mu^+\nu_{\mu})=1.589^{+0.313}_{-0.281}\times 10^{-3}.
\end{align}

\begin{figure*}[t]
\begin{center}
\includegraphics[width=0.4\textwidth]{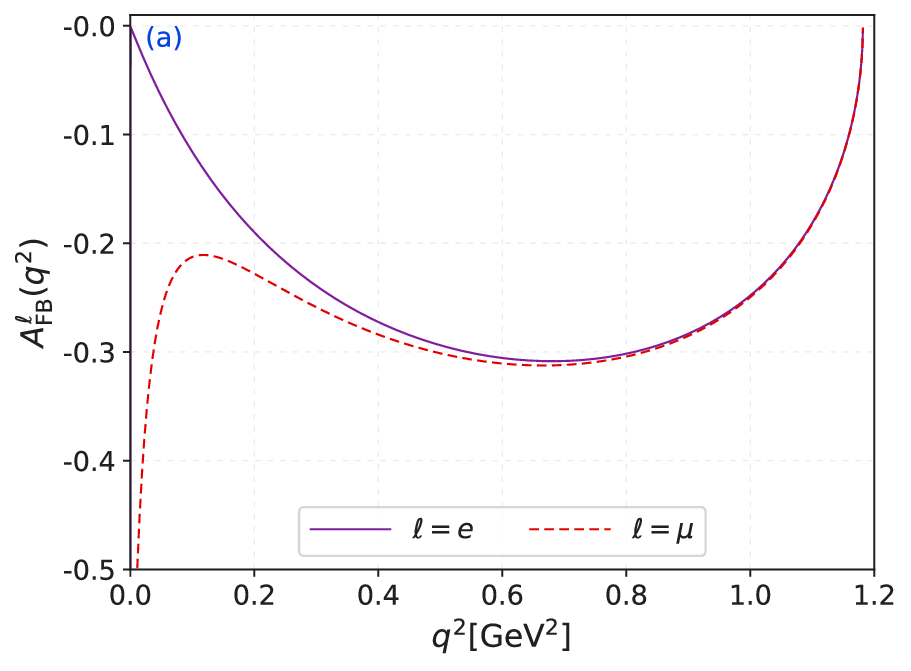}\includegraphics[width=0.4\textwidth]{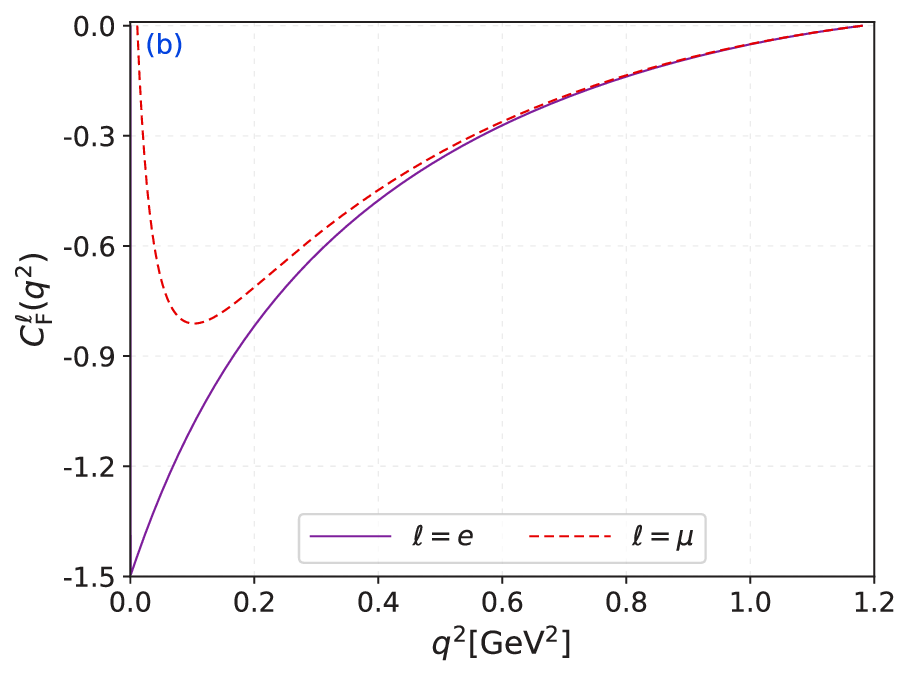}\\
\includegraphics[width=0.4\textwidth]{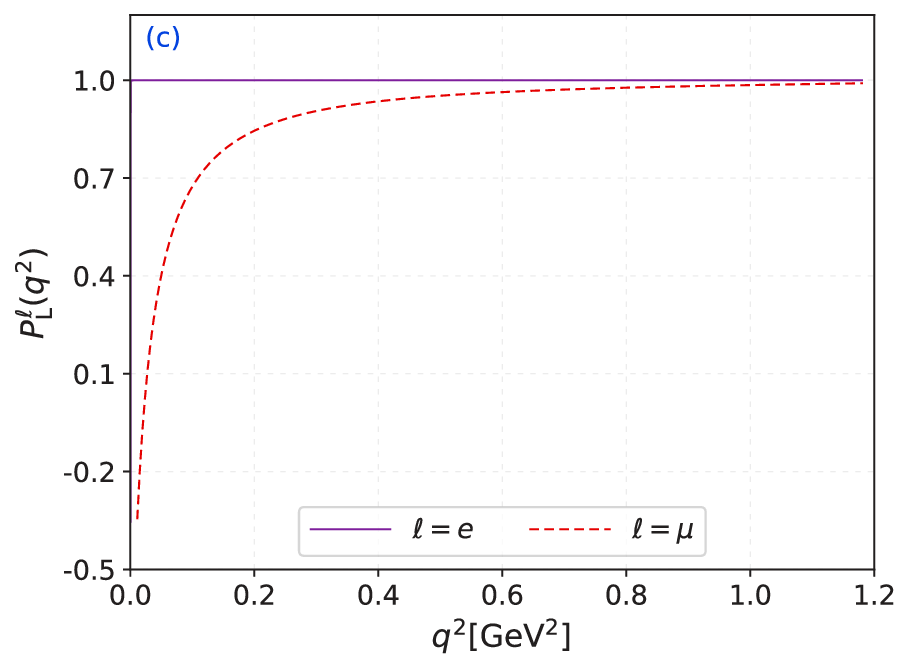}\includegraphics[width=0.4\textwidth]{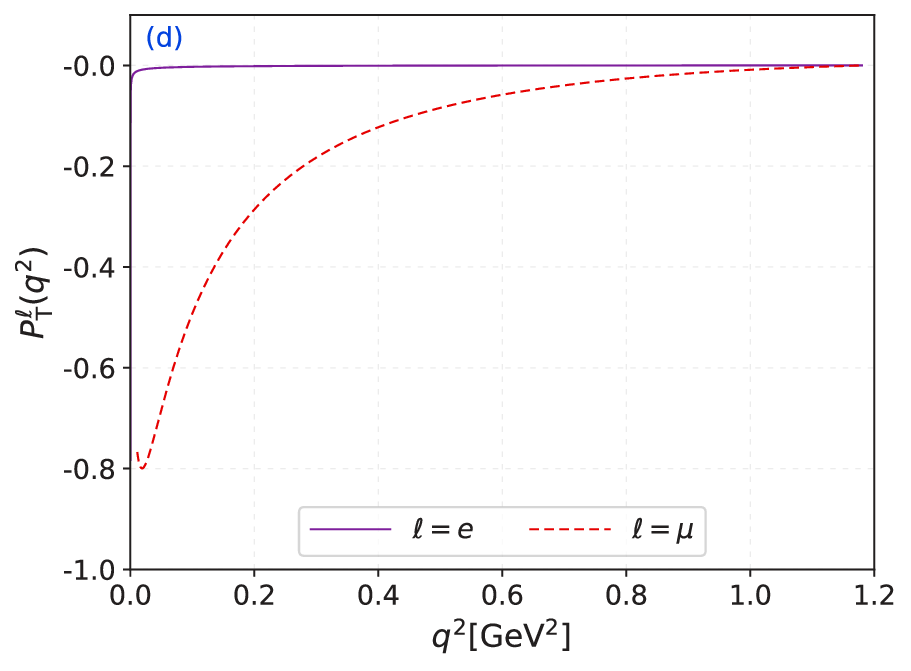}\\
\includegraphics[width=0.4\textwidth]{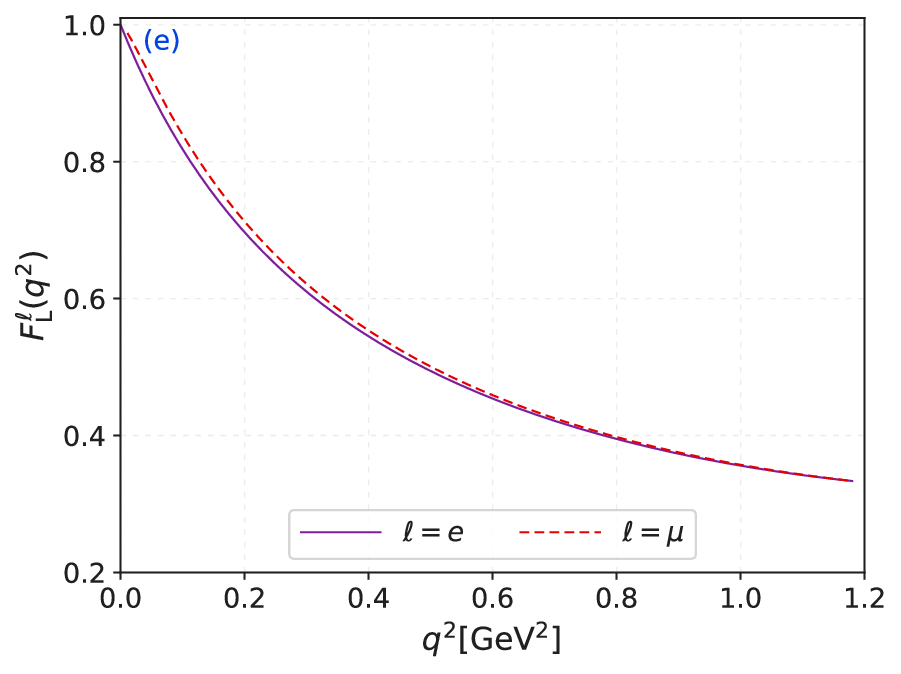}
\end{center}
\caption{The polarization and asymmetry observables as a function of $q^2$, where (a), (b), (c), (d), and (e) denotes forward-backward asymmetry $A_{\rm{FB}}^{\ell}$, lepton-side convexity parameter $C^{\ell}_{\rm{F}}$, longitudinal polarization $P_{\rm{L}}^{\ell}$, transverse polarization $P_{\rm{T}}^{\ell}$ and longitudinal polarization fraction $F_{\rm{L}}^{\ell}$, respectively.}
\label{Fig:Polarization}
\end{figure*}

Finally, we calculate the forward-backward asymmetry $A_{\rm{FB}}^{\ell}$, lepton-side convexity parameter $C^{\ell}_{\rm{F}}$, longitudinal (transverse) polarization $P_{\rm{L}(\rm{T})}^{\ell}$, as well as longitudinal polarization fraction $F_{\rm{L}}^{\ell}$ using Eqs.~\eqref{Eq:AFB}-\eqref{Eq:FL}, which are presented in Fig.~\ref{Fig:Polarization}.
\begin{itemize}
    \item The Fig.~\ref{Fig:Polarization} ({\color{blue}{a}}) show the change of forward-backward asymmetry $A_{\rm{FB}}^{\ell}$ in the range of $m^2_{\ell}< q^2 < (m_{D^+}-m_{\omega})^2$. It can be seen that the curve change of $A_{\rm{FB}}^{\mu}$ in this process is very sharp in the small $q^2$ region, and the two different lepton channels almost have the same result near $q^2 \approx 0.7$, which is $A_{\rm{FB}}^{e/\mu}\approx-0.3$. In the latter part of the region, there is no difference between the two curve trends.
    \item As shown in Fig.~\ref{Fig:Polarization} ({\color{blue}{b}}), the lepton-side convexity parameter $C^{\ell}_{\rm{F}}$ also exhibits a sharp variation in the small $q^2$ region, where the lepton mass effects are most pronounced.At higher $q^2$ region, the $C^{\ell}_{\rm{F}}$ rises smoothly toward zero.
    \item The longitudinal (transverse) lepton polarization $P_{\rm{L}(\rm{T})}^{\ell}$ are presented in Fig.~\ref{Fig:Polarization} ({\color{blue}{c}})-({\color{blue}{d}}). In the whole $q^2$ region, the curve of $P_{\rm{L}(\rm{T})}^{e}$ does not change, and always maintains the characteristics of $P_{\rm{L}}^{e}=1$ and $P_{\rm{T}}^{e}=0$. This indicates that in the limit of lepton mass $m_{\ell}\to 0$, the lepton is purely longitudinally polarized. Compare to $A_{\rm{FB}}^{\ell}$ and $C^{\ell}_{\rm{F}}$, the influence of lepton mass differences on $P_{\rm{L}(\rm{T})}^{\ell}$ is more pronounced across the entire $q^2$ region. At $q^2_{\rm{min}}=m_{\ell}^2$, we observe $P_{\rm{L}}^{\mu}=-0.34$ and $P_{\rm{T}}^{\mu}=-0.77$. As $q^2$ increases, $P_{\rm{L}}^{\mu} $ and $P_{\rm{T}}^{\mu}$ gradually approach 1 and 0, respectively.
    \item For longitudinal polarization fraction $F_{\rm{L}}^{\ell}$ in Fig.~\ref{Fig:Polarization} ({\color{blue}{e}}), the curves of the two lepton channels are almost identical, indicating that the influence of lepton mass is very small. Furthermore, the longitudinal and transverse polarization fractions satisfy $F_{\rm{L}}^{\ell}+F_{\rm{T}}^{\ell}=1$. It can be inferred that at $q^2_{\rm{imn}}=m_{\ell}^2$, $F_{\rm{L}}^{\ell}=1$ necessarily implies $F_{\rm{T}}^{\ell}=0$. In the whole $q^2$ region, the trend of $F_{\rm{L}}^{e/\mu}$ curve is consistent with a decreasing trend, and the corresponding $F_{\rm{T}}^{e/\mu}$ will be an increasing trend.

\end{itemize}
All the above physical observables are expressed as ratio function of TFFs, so the impact brought by the uncertainty of TFF is very small.

\begin{table}[t]
\begin{center}
\renewcommand{\arraystretch}{1.5}
\caption{The mean values of various polarization and asymmetry observables.}
\label{Tab:Psvalue}
\begin{tabular}{l@{\hspace{2em}} c @{\hspace{2em}}c @{\hspace{2em}}c @{\hspace{2em}}c @{\hspace{2em}}c }
\hline
 &$\langle A_\mathrm{FB}^e\rangle$ &$\langle A_\mathrm{FB}^\mu\rangle$ &$\langle C_\mathrm{F}^e\rangle$ &$\langle C_\mathrm{F}^\mu\rangle$ &$\langle P_\mathrm{L}^e\rangle$
\\
\hline
This work     &$-0.23$ &$-0.26$ &$-0.41$ &$-0.32$&$1.00$\\
CCQM~\cite{Ivanov:2019nqd}   &$-0.21$   &$-0.24$ &$-0.43$  &$-0.35$ &$1.00$ \\
RQM~\cite{Faustov:2019mqr}   &$-0.25$   &$-0.27$  &$-0.39$  &$-0.32$  &$1.00$
\\
\hline
&$\langle P_\mathrm{L}^\mu\rangle $ &$\langle P_\mathrm{T}^e\rangle$ &$\langle P_\mathrm{T}^\mu\rangle$ &$\langle F_\mathrm{L}^e\rangle$ &$\langle F_\mathrm{L}^\mu\rangle$
\\
This work &$0.89$ &$0.00$ &$-0.14$ &$0.52$ &$0.51$\\
CCQM~\cite{Ivanov:2019nqd}   &$0.92$   &$0.00$ &$-0.12$  &$0.52$ &$0.50$ \\
RQM~\cite{Faustov:2019mqr}   &$0.93$     &$0.00$   &$-0.11$   &$0.51$  &$0.50$
\\
\hline
\label{table:polarization}
\end{tabular}
\end{center}
\end{table}

\section{Summary}\label{sec:IV}
In this paper, we first compute the four TFFs of $D^+\to \omega$ by employing the right-handed chiral current within the framework of LCSR. For the transverse twist-2 LCDA $\phi^{\perp}_{2;\omega}(x,\mu)$ that dominates the contribution in TFFs, we construct a LCHO model based on the BHL prescription. The Fig.~\ref{Fig:DA} shows that our prediction is similar with other theoretical groups and has good ending-point behavior. Furthermore, we present predictions of $A_{1,2,0}(0)$ and $V(0)$ at $q^2=0$ in Table~\ref{table:lparge point}, which show good agreement with results from other theoretical results.

After extrapolating the TFFs to the higher $q^2$ region using the SSE method, our results show good consistency with most theoretical groups in Fig.~\ref{Fig:TFFscomparison}. The extrapolation trend exhibits a stable behavior across the entire $q^2$ range. Then, we utilize TFFs to calculate the differential decay widths, branching fractions and polarization and asymmetry observables. In Fig.~\ref{Fig:dGamma}, we present the differential decay width $1/|V_{cd}|^2d\Gamma(D^+\to \omega\ell^+\nu_{\ell})/dq^2$. For both the electron and muon channels, we present the corresponding branching fractions in Table~\ref{table:Br}. The result for $\mathcal{B}(D^+\to \omega e^+\nu_e)$ shows good agreement with those from CLEO'11~\cite{CLEO:2011ab}, CCQM~\cite{Soni:2018adu}, while $\mathcal{B}(D^+\to \omega \mu^+\nu_{\mu})$ is consistent with values reported by PDG~\cite{ParticleDataGroup:2024cfk}, BESIII'20~\cite{Ablikim:2020tmg}, CCQM~\cite{Soni:2018adu}, and LCSR~\cite{Fu:2020vqd}. Taking advantage of the narrow width of $\omega$-meson, we also predict the branching fraction for five body decay $D^+\to \omega(\to \pi^+\pi^-\pi^0)\ell^+\nu_{\ell}$. Finally, we discuss the forward-backward asymmetry $A_{\rm{FB}}^{\ell}$, lepton-side convexity parameter $C^{\ell}_{\rm{F}}$, longitudinal (transverse) polarization $P_{\rm{L}(\rm{T})}^{\ell}$, as well as longitudinal polarization fraction $F_{L}^{\ell}$. The trends of their changes with $q^2$ are presented in Fig.~\ref{Fig:Polarization}, and the corresponding mean values are listed in Table~\ref{table:polarization}, showing consistency with predictions from CCQM~\cite{Ivanov:2019nqd} and RQM~\cite{Faustov:2019mqr}.

It is believed that in the near future, this process will be further investigated by the BESIII Collaboration. With the current data sample being larger than previous ones, there is significant potential for the experimental measurement of polarization and asymmetry observables in $D^+\to \omega\ell^+\nu_{\ell}$. The predictions of the physical observables given in our work can not only provide a reference value for experiment collaboration, but also reversely test our LCHO model of twist-2 LCDA and TFFs.

\acknowledgments

This work was supported in part by the National Natural Science Foundation of China under Grant No.12265010 and No.12265011, the Project of Guizhou Provincial Department of Science and Technology under Grant No.MS[2025]219, No.CXTD[2025]030, No.YQK[2023]016 and ZK[2023]141.

\end{document}